\documentclass[journal=mamobx,layout=twocolumn]{achemso}

\usepackage{amsmath,amssymb}
\usepackage[T1]{fontenc}       % Use modern font encodings
\usepackage{graphicx}
\usepackage{braket}            % Use for the average (<...>) symbol.
\usepackage{booktabs}          % Use for nice tables.
\usepackage{color}
\usepackage[dvipsnames]{xcolor}
\usepackage{placeins}
\usepackage[normalem]{ulem}             % Make strikout (\sout) available
\usepackage{float}
\usepackage{graphicx,wrapfig,lipsum}

\usepackage{graphicx}% Include figure files
\usepackage{dcolumn}% Align table columns on decimal point
\usepackage{bm}% bold math
\usepackage{braket}            % Use for the average (<...>) symbol.
\usepackage{color}
\usepackage[dvipsnames]{xcolor}
\usepackage{placeins}
\usepackage{float}

\renewcommand{\vec}[1]{\mathbf{#1}} % Vectors in bold.
\newcommand{\uvec}[1]{\mathbf{\widehat{#1}}} % Unitvectors in bold.
\newcommand{\dd}{\mathrm{d}}
\newcommand\norm[1]{\left\lVert#1\right\rVert}
\newcommand{\ra}[0]{\right\rangle}
\newcommand{\la}[0]{\left\langle}

\newcommand{\DID}[1]{^{(#1)}}

\title{DNA elasticity from coarse-grained simulations:
the effect of groove asymmetry}

\author{Enrico Skoruppa}
\affiliation{KU Leuven, Institute for Theoretical Physics,
Celestijnenlaan 200D, 3001 Leuven, Belgium}
\author{Michiel Laleman}
\affiliation{KU Leuven, Institute for Theoretical Physics,
Celestijnenlaan 200D, 3001 Leuven, Belgium}
\author{Stefanos K.\ Nomidis}
\affiliation{KU Leuven, Institute for Theoretical Physics,
Celestijnenlaan 200D, 3001 Leuven, Belgium}
\affiliation{Flemish Institute for Technological Research (VITO), Boeretang
200, B-2400 Mol, Belgium}
\author{Enrico Carlon}
\affiliation{KU Leuven, Institute for Theoretical Physics,
Celestijnenlaan 200D, 3001 Leuven, Belgium}
\email{enrico.carlon@kuleuven.be}

\date{\today}% It is always \today, today,
\keywords{DNA, oxDNA, twist-bend coupling}%Use showkeys class option if keyword

\begin{document}

\begin{abstract}
It is well-established that many physical properties of DNA at
sufficiently long length scales can be understood by means of simple
polymer models. One of the most widely used elasticity models for DNA
is the twistable worm-like chain (TWLC), which describes the double
helix as a continuous elastic rod with bending and torsional stiffness.
An extension of the TWLC, which has recently received some attention,
is the model by Marko and Siggia, who introduced an additional
twist-bend coupling, expected to arise from the groove asymmetry. By
performing computer simulations of two available versions of oxDNA,
a coarse-grained model of nucleic acids, we investigate the microscopic
origin of twist-bend coupling. We show that this interaction is negligible
in the oxDNA version with symmetric grooves, while it appears in the
oxDNA version with asymmetric grooves. Our analysis is based on the
calculation of the covariance matrix of equilibrium deformations,
from which the stiffness parameters are obtained. The estimated
twist-bend coupling coefficient from oxDNA simulations is $G=30\pm1$~nm.
The groove asymmetry induces a novel twist length scale and an associated
renormalized twist stiffness $\kappa_{\rm t} \approx 80$~nm, which is
different from the intrinsic torsional stiffness $C \approx 110$~nm.
This naturally explains the large variations on experimental estimates
of the intrinsic stiffness performed in the past.
\end{abstract}

%  \pacs{Valid PACS appear here}% PACS, the Physics and Astronomy
\maketitle


\section{Introduction}

Owing to its role as the carrier of genetic information, DNA is of central
importance in biology. In its interactions with other biomolecules within
the cell, DNA is often bent and twisted.  A good mechanical model of DNA
is therefore essential to understand the complex biological processes
in which it is involved.~\cite{brya12}  A large number of experiments
in the past have shown that its mechanical response can be described
using simple continuous polymer models (studies of such models can be
found e.g.\ in Refs.~\citenum{mark95, moro98, ubbi99}), such as the
twistable worm-like chain (TWLC), which treats DNA as an elastic rod,
exhibiting resistance to applied bending and twisting~\cite{nels08}. In
spite of its simplicity, the TWLC has proven to be surprisingly accurate
in the description of the DNA response to applied forces~\cite{mark95,
bust03} and torques~\cite{mark94b, stri96}.

As experimental techniques become more accurate, physical models
are put to increasingly strict tests. Single-molecule experiments of
the past few years have reported some discrepancies between the TWLC
predictions and the observed torsional response of DNA \cite{lipf10,
lipf11}. These experiments use magnetic tweezers in order to apply
both a torque and a stretching force to a single DNA molecule. The
measured torsional stiffness as a function of the applied force turned
out to deviate from the TWLC predictions. A recent study explained
these discrepancies using an elastic DNA model, which extends the TWLC
by including a direct coupling term between the twisting and bending
degrees of freedom~\cite{nomi17}. The existence of twist-bend coupling
was already predicted by Marko and Siggia~\cite{mark94} in 1994.
Quite surprisingly the consequence of this coupling on the structural
and dynamical properties of DNA has only been discussed in a very limited
number of papers so far~\cite{lank00,moha05}.

In this paper we investigate the elastic properties of oxDNA, a
coarse-grained model for simulations of single- and double-stranded
DNA~\cite{ould10}. OxDNA comes in two versions: the original
version (oxDNA1) contains symmetric grooves, while in a more
recent extension (oxDNA2) distinct major and minor grooves were
introduced~\cite{snod15}. By comparing the two versions, we deduce
the effect of an asymmetric grooving on the elastic properties of the
molecule.  Our analysis shows a clear signature of twist-bend coupling
in oxDNA2, while this interaction is absent in the symmetric oxDNA1.
This confirms the predictions of Marko and Siggia~\cite{mark94} and
shows that the groove asymmetry strongly affects the elastic properties
of the molecule.  Our estimate of the twist-bend coupling constant in
oxDNA2 is in agreement with that obtained from a recent analysis of
magnetic tweezers data~\cite{nomi17}.

\section{Models and simulations}

\subsection{Elasticity models}

\begin{figure}[t]
\centering\includegraphics{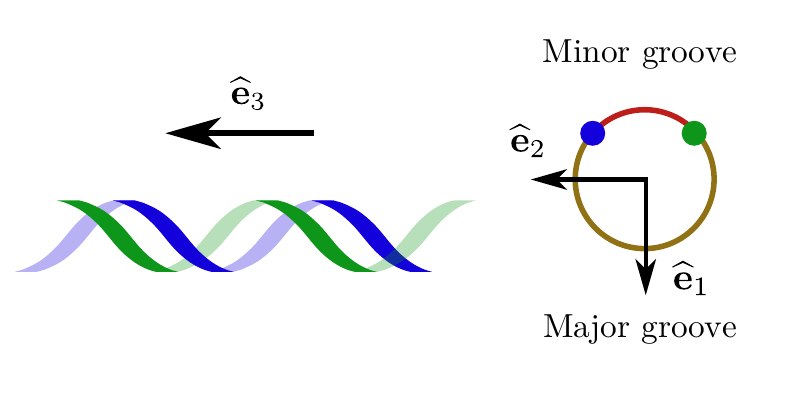}
\caption{DNA can be represented as an inextensible, twistable, elastic
rod. Its conformation is described by a local orthonormal frame,
associated with every point along the molecule. $\vec{\widehat e}_3$
is the unit tangent vector, whereas $\vec{\widehat e}_1$ is chosen to
lie on the symmetry plane of the grooves. The third vector is given by
$\vec{\widehat{e}}_2 = \vec{\widehat{e}}_3 \times \vec{\widehat{e}}_1$.}
\label{fig:frame}
\end{figure}

Elastic polymer models describe double-stranded DNA as a continuous
inextensible rod. At every point along the molecule one defines a local
frame of reference, given by a set of three orthonormal vectors $\{
\vec{\widehat{e}}_1(s), \vec{\widehat{e}}_2(s), \vec{\widehat{e}}_3(s)
\}$, where $0 \le s \le L$ is the arc-length coordinate and $L$ the
contour length. The common convention is to choose $\vec{\widehat{e}}_3$
as local tangent to the curve (see Fig.~\ref{fig:frame}), whereas
$\vec{\widehat{e}}_1$ and $\vec{\widehat{e}}_2$ lie in the plane
of the ideal, planar Watson-Crick base pairs~\cite{mark94}. The
vector $\vec{\widehat{e}}_1$ is directed along the symmetry axis
of the two grooves and $\vec{\widehat{e}}_2$ is obtained from
the relation $\vec{\widehat{e}}_2 = \vec{\widehat{e}}_3 \times
\vec{\widehat{e}}_1$. Knowing how the set $\{ \vec{\widehat{e}}_1(s),
\vec{\widehat{e}}_2(s), \vec{\widehat{e}}_3(s) \}$ depends on $s$ allows
one to reconstruct the conformation of the molecule.

%%%%%%%%%%%%%%%%%%%%%%%%%%%%%%%%%%%%%%%%%%%%%%%%%%%%%%%%%%%%%%%%%%%%%%%%%%%
%%%%%%%%%%%%%%%%%%%%%%%%%%%%%%%%%%%%%%%%%%%%%%%%%%%%%%%%%%%%%%%%%%%%%%%%%%%

Any local deformation of the curve induces a rotation of the frame
$\{ \vec{\widehat{e}}_1, \vec{\widehat{e}}_2, \vec{\widehat{e}}_3 \}$
from $s$ to $s+ds$, which can be described by the following 
differential equation
\begin{equation}\label{eq:rotation}
\frac{\dd\vec{\widehat{e}}_\mu}{\dd s} = 
( \vec{\Omega} + \omega_0 \vec{\widehat{e}}_3 ) 
\times \vec{\widehat{e}}_\mu,
\end{equation}
where $\mu = 1,2,3$, and $\omega_0$ is the intrinsic twist density of the
DNA double helix. The vector $\vec{\Omega} + \omega_0 \vec{\widehat{e}}_3$
is parallel to the axis of rotation from $\vec{\widehat{e}}_\mu(s)$ to
$\vec{\widehat{e}}_\mu(s+ds)$.  Note that in general $\vec{\Omega}(s)$
depends on the coordinate $s$.  Decomposing this vector along the
local frame, we define its three components as $\Omega_\mu (s) \equiv
\vec{\Omega} \cdot \vec{\widehat{e}}_\mu(s)$.  The case $\vec{\Omega}
= |\vec{\Omega}|\vec{\widehat{e}}_3$ corresponds to a pure twist
deformation, whereas $\vec{\Omega} = |\vec{\Omega}|\vec{\widehat{e}}_1$
and $\vec{\Omega} = |\vec{\Omega}|\vec{\widehat{e}}_2$ express bending
in the planes defined by $\vec{\widehat{e}}_1$ and $\vec{\widehat{e}}_2$,
respectively.

The lowest-energy configuration of the system is that of zero mechanical
stress $\Omega_1 = \Omega_2 = \Omega_3 = 0$, which corresponds to a
straight rod with an intrinsic twist angle per unit length equal to $\omega_0$.  Expanding around this ground state, one obtains the elastic
energy to lowest order in the deformation parameters $\Omega_\mu$ as
\begin{equation}\label{eq:energy_gen}
\beta E = \frac{1}{2}\int_0^L \sum_{\mu,\nu=1}^3
\Omega_\mu (s) M_{\mu\nu} \Omega_\nu(s) ds,
\end{equation}
where $\beta \equiv 1/k_\mathrm{B}T$ is the inverse temperature. The $3\times3$
symmetric matrix $M_{\mu\nu}$, which we refer to as the stiffness matrix,
contains the elastic constants. Note that from Eq.~(\ref{eq:rotation}) the
$\Omega$'s have the dimension of inverse length. As the left-hand side of
Eq.~(\ref{eq:energy_gen}) is dimensionless, the elements of the stiffness
matrix have the dimension of length. In this work sequence-dependent
effects will be neglected, therefore $\vec{M}$ will not depend on~$s$.

Marko and Siggia~\cite{mark94} argued that, due to the asymmetry
introduced by the major and minor grooves, the elastic energy of
DNA should be invariant only under the transformation $\Omega_1 \to
-\Omega_1$. This implies that $\Omega_2 \Omega_3$ is the only 
cross-term allowed by symmetry, therefore the stiffness matrix in the 
Marko-Siggia
(MS) model becomes
\begin{equation}\label{eq:energy_ms}
 \vec{M}_\text{MS} = \begin{pmatrix}
                A_1 & 0   & 0 \\[4pt]
                0   & A_2 & G \\[4pt]
                0   & G   & C
               \end{pmatrix},
\end{equation}
where $A_1 \equiv M_{11}$, $A_2 \equiv M_{22}$,
$C \equiv M_{33}$ and $G \equiv M_{23} = M_{32}$. $A_1$ and $A_2$
express the energetic cost of a bending deformation about the
local axes $\vec{\hat{e}}_1$ and $\vec{\hat{e}}_2$, respectively~\cite{sala15}. 
$C$
is the intrinsic torsional stiffness, whereas $G$ quantifies the
twist-bend coupling interaction.  Note that $G \neq 0$ is a direct
consequence of the groove asymmetry in the DNA double helix. If one
neglects this asymmetry, the MS model reduces to the TWLC model ($G=0$),
and the corresponding stiffness matrix becomes diagonal~\cite{mark94}
\begin{equation}\label{eq:energy_twlc}
\vec{M}_\text{TWLC} = 
\begin{pmatrix}
                A_1 & 0   & 0 \\[4pt]
                0   & A_2 & 0 \\[4pt]
                0   & 0   & C
\end{pmatrix}.
\end{equation}
Most studies~\cite{nels08} model DNA as an isotropic TWLC, for which
$A_1=A_2$.

\subsection{Computer simulations with oxDNA}

In this paper we investigate the elastic properties of oxDNA, which
is a model for coarse-grained computer simulations of both single- and
double-stranded DNA~\cite{ould10}.  The model describes 
double-stranded DNA as two intertwined strings
of rigid nucleotides, with pairwise interactions modeling the backbone
covalent bonds, the hydrogen bonding, the stacking, cross-stacking and
excluded-volume interactions. oxDNA has been used in the past for the
study of a variety of DNA properties~\cite{ould10,sulc12,snod15,sutt16}.

%%%%%%%%%%%%%%%%%%%%%%%%%%%%%%%%%%%%%%%%%%%%%%%%%%%%%%%%%%%%%%%%%%%%%%%%%%%
\begin{figure}[t]
\centering\includegraphics[width=\linewidth]{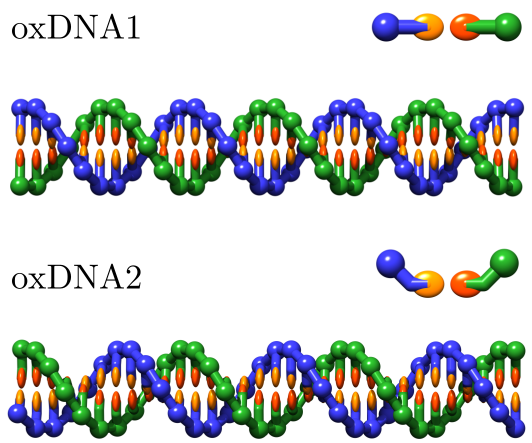}
\caption{Snapshots of configurations of oxDNA1 (top) and oxDNA2
(bottom), including a cross-section view of the helix. While the grooves 
are symmetric in oxDNA1, distinct major and minor grooves are present
in oxDNA2.}
\label{fig:oxDNA1_2}
\end{figure}
%%%%%%%%%%%%%%%%%%%%%%%%%%%%%%%%%%%%%%%%%%%%%%%%%%%%%%%%%%%%%%%%%%%%%%%%%%%

We performed simulations using two available versions of the model. The
first version (oxDNA1) describes DNA as a molecule with no distinction
between major and minor grooves~\cite{sulc12}, while the second
(oxDNA2) introduces distinct grooving asymmetry~\cite{snod15}.
Figure~\ref{fig:oxDNA1_2} illustrates molecular conformations of the
two models, including a cross-sectional view of a single base pair. As
discussed above, the presence of distinct major and minor grooves
breaks a molecular symmetry, so we expect that oxDNA1 and oxDNA2 will
be mapped onto the TWLC (Eq.~(\ref{eq:energy_twlc})) and the MS model
(Eq.~(\ref{eq:energy_ms})), respectively.

To sample equilibrium fluctuations, molecular dynamics simulations in
the NVE ensemble with an Anderson-like thermostat were used. This is
implemented in repeated cycles in which the system is first evolved by
integrating Newton's equations of motion in time for a given number of
steps. Then the momenta of some randomly selected particles are chosen
from a Maxwell distribution with a desired simulation temperature
($T=295$~K in our case). The cycle then repeats itself a large number
of times.

Molecular dynamics simulations were performed on 150 base pair
molecules using averaged base pair interaction coefficients. A total of
$5\times10^{10}$ time steps were sampled using a numerical integration
time step of $15.2$ fs, and the trajectories were recorded every
$5\times10^4$ time steps. For all simulations the salt concentration was
set to $0.5$~M. In oxDNA1 this value is fixed, since the electrostatic
interactions are implemented through excluded-volume potentials,
parametrized to mimic high salt concentration (i.e.\ $0.5$~M). oxDNA2
improved upon this approach by switching to a Debye-H\"{u}ckel potential,
which models the ionic screening of electrostatic interactions.  This
allows for the explicit selection of a salt concentration, which we set to
$0.5$~M, in order to achieve optimal comparability between the two models.

\subsection{Extraction of elastic parameters}

The pivotal objective of the extraction of elastic parameters is
to map oxDNA onto the described elastic model in such a way, that
both the elastic properties at the base pair level as well as long
range behavior, such as bending and torsional persistence lengths,
are captured as accurately as possible. Establishing an appropriate
one-to-one correspondence requires the reduction of both models to
the same level of complexity. For the continuous elastic model this
implies the discretization of the elastic free energy functional
Eq.~(\ref{eq:energy_gen}) to the base pair level
\begin{eqnarray}
\beta E &=& \frac{a}{2} \sum_{n=1}^N 
\left(\sum_{\mu,\nu=1}^3 
\Omega_\mu\DID{n}
M_{\mu\nu}
\Omega_\nu\DID{n}\right),
\label{eq:elastic_energy_discrete}
\end{eqnarray}
where $a = 0.34$~nm is the mean distance between successive base pairs and
$\Omega_\mu\DID{n}\equiv \Omega_\mu (na)$. In the discrete case the finite
rotation of a local frame of reference (triad) $\{ \vec{\widehat{e}}_1(n),
\vec{\widehat{e}}_2(n), \vec{\widehat{e}}_3(n) \}$, associated with
the spatial orientation of the $n$-th base pair of the molecule,
into the sequentially adjacent triad $\{ \vec{\widehat{e}}_1(n+1),
\vec{\widehat{e}}_2(n+1), \vec{\widehat{e}}_3(n+1) \}$, can be
represented by a rotation vector $\vec{\Theta}\DID{n}$. The deformation
parameters $\Omega_\mu\DID n$ can then be defined as the deviations of
the components of $\vec{\Theta}\DID{n}/a$ from their respective mean
values \begin{equation} a\Omega_\mu\DID{n} \equiv \Theta_\mu\DID{n} -
\la \Theta_\mu\DID{n} \ra.  \end{equation} For oxDNA1 the mean twist angle
$a\omega_0 = \langle \Theta_3\DID n\rangle $ is found to be $34.8^\circ$,
whereas for oxDNA2 we find $34.1^\circ$.

Accordingly, an appropriate triad has to be assigned to each base pair
of the oxDNA model.  The particular choice of those triads contains a
certain degree of ambiguity, resulting in different mappings for different
triads. Such an ambiguity regarding the definition of the tangent vector
$\vec{\widehat{e}}_3$ in coarse-grained simulations of DNA and the related
implications for the extraction of the bending persistence length have
for instance been discussed by Fathizadeh et al.~\cite{fath12}, who
showed that, when considering short length scales, different definitions
of the local tangent vector will usually yield significantly different
results for the bending persistence length.  However, when considering
longer length scales, i.e.\ comparing more distant tangent vectors,
those discrepancies vanish asymptotically.\newline For a detailed
discussion of different triad definitions we refer to the Supplementary
Material. All results presented in the main text are calculated with
a triad definition employing local tangents $\vec{\widehat{e}}_3$
obtained from the mean vector of the intrinsic orientation of the two
nucleotides in each basepair, provided by the oxDNA output. The unit
vector $\vec{\widehat{e}}_2$ is obtained from the projection of the
connecting vector between the centers of the two nucleotides $\vec{y}$,
onto the orthogonal space of $\vec{\widehat{e}}_3$.  Having identified
$\vec{\widehat{e}}_3$ and $\vec{\widehat{e}}_2$ the remaining vector in
the right-handed triad is now uniquely defined as
 $\vec{\widehat{e}}_1 = \vec{\widehat{e}}_2 \times
\vec{\widehat{e}}_3$. This corresponds to Triad II in the Supplementary
Material.

%%%%%%%%%%%%%%%%%%%%%%%%%%%%%%%%%%%%%%%%%%%%%%%%%%%%%%%%%%%%%%%%%%%%%%%%%%%%%%
\begin{figure*}[t]
\centering\includegraphics[width=\linewidth]{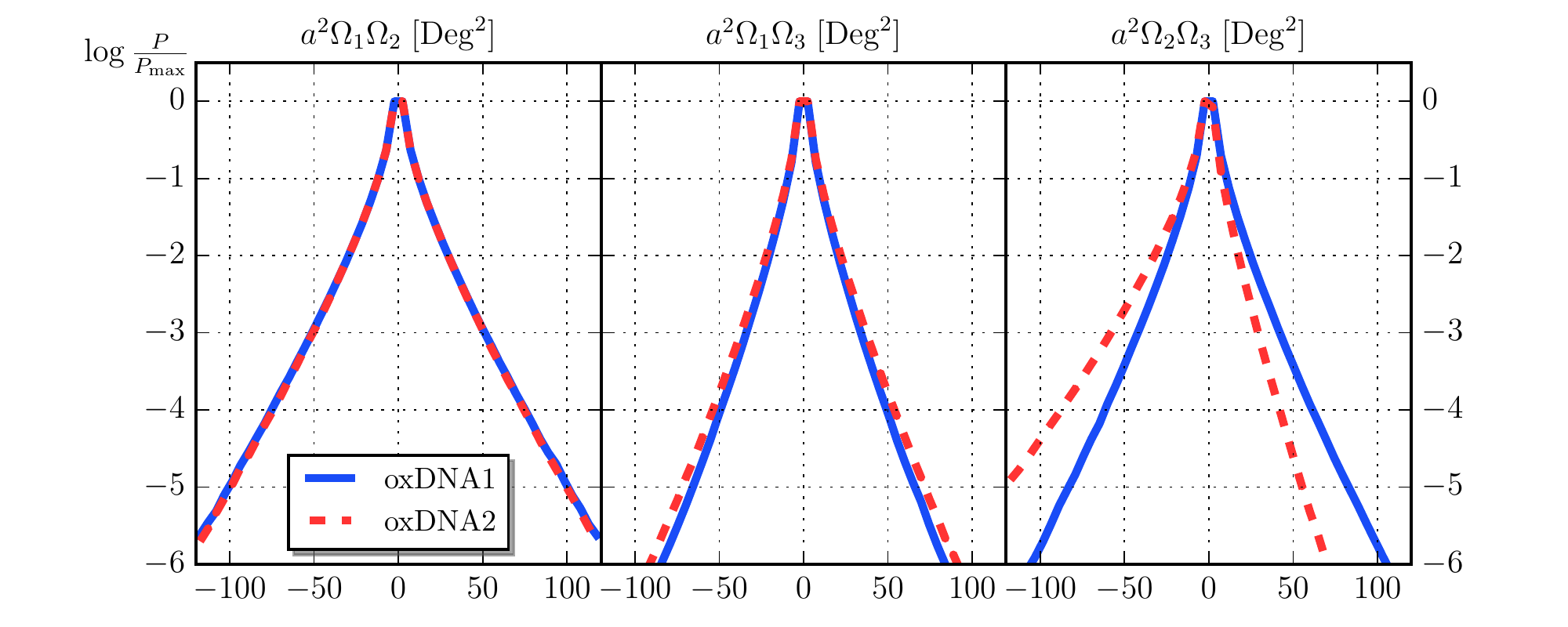}
\caption{Histograms of cross-diagonal terms $\Omega_\mu \Omega_\nu$ for
oxDNA1 and oxDNA2. The histograms for $\Omega_1 \Omega_2$ and $\Omega_1
\Omega_3$ are quite similar for the two models, while there is a marked
difference for $\Omega_2 \Omega_3$. The asymmetric shape of the histogram
in oxDNA2 is a signature of the presence of twist-bend coupling.}
\label{fig:hist_cross}
\end{figure*}
%%%%%%%%%%%%%%%%%%%%%%%%%%%%%%%%%%%%%%%%%%%%%%%%%%%%%%%%%%%%%%%%%%%%%%%%%%%%%%

In order to infer the stiffness matrix from simulations, we used the
standard procedure (see e.g.\ Ref.~\citenum{lank00}) which relies on
the equipartition theorem~\cite{huang}
\begin{equation}
\left\langle \Omega_\mu\DID{n}  \, 
\frac{\partial \beta E}{\partial \Omega_\nu\DID{n} } 
\right\rangle = \delta_{\mu\nu},
\label{eq:EPT}
\end{equation}
where $\langle \cdot \rangle$ indicates the thermal average.  
Then we introduced the $3 \times 3$ covariance matrix with elements
\begin{equation}
\Lambda_{\mu \nu} \equiv 
\left\langle \Omega_\mu\DID{n} \,\Omega_\nu\DID{n} \right\rangle,
\label{def_lambda}
\end{equation}
where the index $n$ was dropped from $\Lambda$, as we neglect
sequence-dependent effects. Combining (\ref{eq:elastic_energy_discrete})
and (\ref{eq:EPT}) we get
\begin{equation} \vec{M} =
\frac{1}{a}\vec{\Lambda}^{-1}. 
\label{eq:invStiff} 
\end{equation}
Thus, the stiffness parameters contained in $\vec{M}$ can be extracted
from the correlation matrix $\vec{\Lambda}$, obtained from equilibrium
fluctuations (Eq.~(\ref{def_lambda})).

This procedure is based on the elastic energy being given by
Eq.~(\ref{eq:elastic_energy_discrete}), which in turn assumes that there
are no correlations between different sets of $\Omega$'s.  To investigate
the effect of correlations we introduce the matrix
\begin{equation}
\Xi_{\mu\nu} (m) \equiv 
\left\langle 
\left[
\sum_{k=n}^{n+m-1} 
\Omega_\mu\DID{k} 
\right]
\left[
\sum_{l=n}^{n+m-1} 
\Omega_\nu\DID{l} 
\right]
\right\rangle
\label{eq:Gammam}.
\end{equation}
If correlations beyond neighboring bases are weak, the cross-terms in the 
previous expression can be neglected and we obtain
\begin{equation}
\Xi_{\mu\nu} (m) \approx \sum_{k=n}^{n+m-1} \left\langle 
\Omega_\mu\DID{k} \Omega_\nu\DID{k}
\right\rangle
=m \Lambda_{\mu\nu}.
\end{equation}
Finally we define the $m$-step stiffness matrix as
\begin{equation}
\vec{M}(m) \equiv \frac{m}{a} \left[ \vec{\Xi}(m) \right]^{-1},
\label{eq:Mm}
\end{equation}
from which the $m$-step elastic constants can be obtained. In absence of
correlations, this matrix will not depend on $m$.

%%%%%%%%%%%%%%%%%%%%%%%%%%%%%%%%%%%%%%%%%%%%%%%%%%%%%%%%%%%%%%%%%%%
\begin{figure*}[t]
\centering\includegraphics[width=\linewidth]{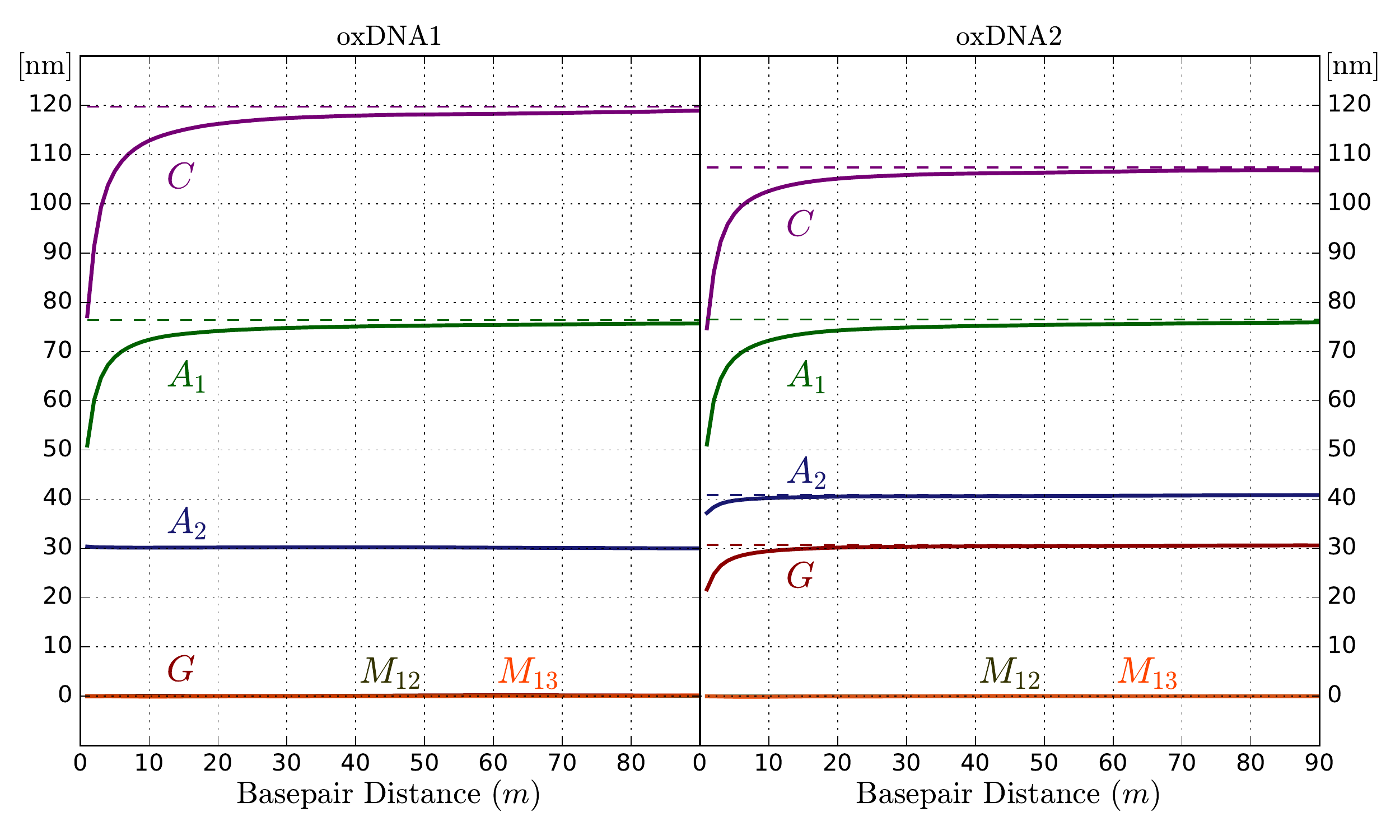}
\caption{
Elastic parameters, obtained from the $m$-step stiffness matrix,
as a function of the base pair distance $m$. The remarkable difference
between these two sets is the appearance of a significant twist-bend
coupling term $G$ for oxDNA2, in contrast to its negligible value in
oxDNA1. This is in agreement with the original prediction of Marko and
Siggia~\cite{mark94}.}
\label{fig:elast_param}
\end{figure*}
%%%%%%%%%%%%%%%%%%%%%%%%%%%%%%%%%%%%%%%%%%%%%%%%%%%%%%%%%%%%%%%%%%%

\section{Results}

We present here the results of the simulations highlighting the
differences in elastic properties between oxDNA1 and oxDNA2.

\paragraph{\textbf{Probability Distributions}}

Qualitative evidence of the presence of a non-zero twist-bend coupling
in the energy functionals can already be inferred from the distribution
of the off-diagonal terms $\Omega_\mu\DID{n} \Omega_\nu\DID{n}$ with
$\mu \neq \nu$.  Figure~\ref{fig:hist_cross} shows histograms of these
quantities, obtained from simulations of oxDNA1 and oxDNA2. The data are
averaged over all base pairs along the DNA contour, hence we drop the
position index $n$.  While the distribution of $\Omega_1\Omega_2$ and
$\Omega_1\Omega_3$ is symmetric and very similar in oxDNA1 and oxDNA2,
there is a marked difference between the two models in the histogram of
$\Omega_2\Omega_3$. In oxDNA1 the distribution appears to be symmetric,
whereas in oxDNA2 there is a clear asymmetry, suggesting the existence
of a coupling between those deformation parameters.

%%%%%%%%%%%%%%%%%%%%%%%%%%%%%%%%%%%%%%%%%%%%%%%%%%%%%%%%%%%%%%%%%%%
\begin{figure*}[t]
\centering\includegraphics[width=\linewidth]{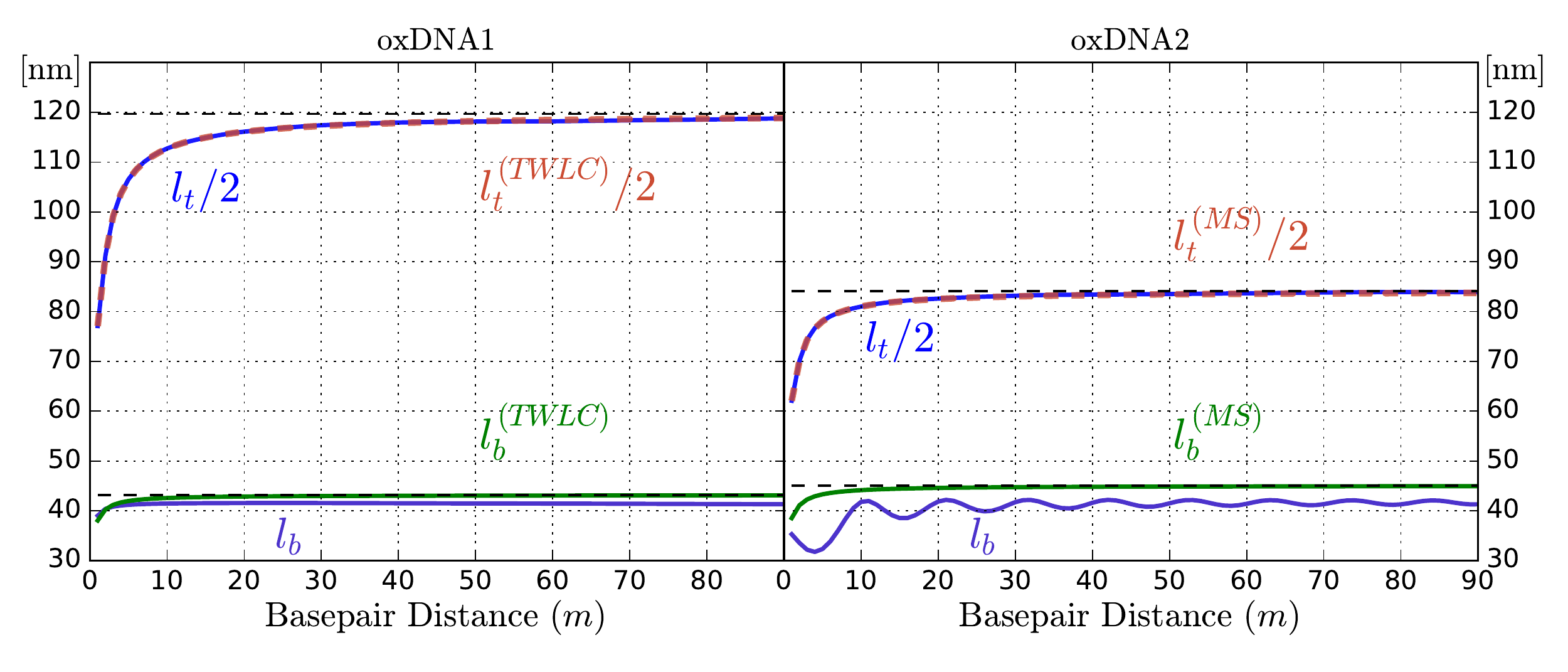}
\caption{
Blue lines: plots of $l_\text{b}$ and $l_\text{t}/2$ obtained
from oxDNA simulations using Eqs.~\eqref{eq:lbm_numerics} and
\eqref{eq:ltm_numerics}. Orange and green lines: analytical predictions
for the same quantities in the TWLC (Eqs.~\eqref{eq:lb_TWLC} and
\eqref{eq:lt_TWLC}) and in the MS model (Eqs.~(\eqref{eq:lb_MS}
and \eqref{eq:lt_MS}), where the $m$-dependent stiffnesses of
Fig.~\ref{fig:elast_param} were used.The values obtained from the plateau
values of the elastic parameters are indicated by the dashed black lines.}
\label{fig:lb-lt}
\end{figure*}
%%%%%%%%%%%%%%%%%%%%%%%%%%%%%%%%%%%%%%%%%%%%%%%%%%%%%%%%%%%%%%%%%%%

\paragraph{\textbf{Stiffness Matrix}}

In order to quantify the observed twist-bend coupling interaction,
we computed the $m$-step stiffness matrix~$\vec{M}(m)$, as defined in
Eq.~\eqref{eq:Mm}, for both models and for different summation lengths
$m$. At both chain-ends 5 base pairs were excluded from this calculation,
since those boundary segments are found to exhibit a significantly
higher flexibility than segments located in the center of the chain.
 The results are shown in Fig.~\ref{fig:elast_param}, where the
elements of $\vec{M}(m)$ are plotted as a function of $m$.  In both
models the diagonal elements $A_1$, $A_2$ and $C$, as defined in
Eqs.~(\ref{eq:energy_ms}) and \eqref{eq:energy_twlc}, have distinct,
non-vanishing values. There is, however, a remarkable difference between
oxDNA1 and oxDNA2 in the values of the off-diagonal elements $G$, $M_{12}$
and $M_{13}$. In particular, all off-diagonal elements in oxDNA1 are
orders of magnitude smaller than the diagonal ones. On the
other hand, although $M_{12}$ and $M_{13}$ remain negligibly small, the
twist-bend coupling $G$ in oxDNA2 becomes comparable in magnitude to the
diagonal terms, which clearly has to be attributed to the asymmetry of
the helical grooves. These results are in line with the predictions of
Marko and Siggia~\cite{mark94} and remain valid regardless of the exact
choice of coordinate systems (see Supplementary Material).

%%%%%%%%%%%%%%%%%%%%%%%%%%%%%%%%%%%%%%%%%%%%%%%%%%%%%%%%%%%%%%%
\setlength{\tabcolsep}{7pt} % Space between columns
\begin{table*}
\caption{Values of the stiffness coefficients for oxDNA1 and 
oxDNA2 obtained in this work (expressed in nm). The last line 
shows the values obtained from 
fitting the MS model to magnetic tweezers data.}
%  \begin{ruledtabular}
\centering
\begin{tabular}{c c c c c}

% \toprule

& $A_1$ & $A_2$ & C & G \\
	
% \midrule\\[-10pt]
\hline\\[-6 pt]

oxDNA1 & 84(14) & 29(2) & 118(1) & 0.1(0.2) \\ 
oxDNA2 & 81(10) & 39(2) & 105(1) & 30(1)\\
Nomidis et al.~\cite{nomi17}& 66 & 46 & 110(5) & 40(10)\\[4pt]

% \bottomrule

\end{tabular}
\label{table1}
%  \end{ruledtabular}
\end{table*}
%%%%%%%%%%%%%%%%%%%%%%%%%%%%%%%%%%%%%%%%%%%%%%%%%%%%%%%%%%%%%%%

As discussed in the previous section, in absence of correlations between
different sets of $\Omega$'s, the elements of $\vec{M}(m)$ are expected
to be independent of $m$. The results of Fig.~\ref{fig:elast_param},
however, show that this is not exactly true, which is a signature of the
influence of correlations between base pairs separated by more than one
nucleotide (though the convergence to a limiting value for increasing $m$
is quite rapid).

When comparing the results among different choices of frames, we find
that, despite the different values for $m=1$, at large $m$ all values
are close to each other (see Supplementary Material). We, thus, consider
these limiting values to be good estimates for the stiffness parameters
of the elastic model, onto which oxDNA is mapped. Table~\ref{table1}
summarizes the estimated values of the elastic parameters, averaged
over the different choices of local frames, where the error bars
reflect the uncertainty from estimates obtained from four different
definitions of frames.  The first two rows in Table~\ref{table1} are
data obtained from oxDNA simulations in this work, while the last row
shows the parametrization obtained from fits of the MS model to magnetic
tweezers data~\cite{nomi17}.  oxDNA2 data for $C$ and $G$ are consistent
with the latter, while some differences are found in $A_1$ and $A_2$.
 It should be noted, however, that the fitting procedure used in
Ref.~\citenum{nomi17} was not very sensitive to the specific choice of
$A_1$ and $A_2$, as other choices fitted the experimental data equally
well. The overall quantitative agreement between the oxDNA2 parameters
and those from this recent study supports the choice of the plateau values
in Fig.~\ref{fig:elast_param} as an estimate for the elastic parameters.

The value obtained for $C$ is in general good agreement with previous
estimates for oxDNA, which were obtained from methods not involving
the calculation of the stiffness matrix.  From two independent
measurements~\cite{ould11,mate15} the value $C=115$~nm was reported
for oxDNA1. In oxDNA2 a fit of torsional stiffness data~\cite{snod15}
gives $C=93-98$~nm, which is slightly lower than our current estimate.

\paragraph{\textbf{Persistence lengths}}

Any twistable polymer model is characterized by two distinct persistence
lengths, related to bending and twisting fluctuations. The bending
persistence length can be obtained from the decay of the correlation
between tangent vectors
\begin{equation}
\left\langle \vec{\widehat{e}}_3 (n) \cdot \vec{\widehat{e}}_3 (n+m) 
\right\rangle \equiv \langle \cos \theta(m) \rangle
\sim e^{-ma/l_\text{b}},
\end{equation}
where $\theta(m)$ is the angle formed by the two vectors. As the 
exponential decay is valid asymptotically in $m$, we can
estimate the bending persistence length from the extrapolation at large
$m$ of the quantity 
\begin{equation}
l_\text{b} (m) \equiv -\frac{ma}{\log \left\langle \cos \theta(m) 
\right\rangle}.
\label{eq:lbm_numerics}
\end{equation}
Analogously, we can define the twisting persistence length from the
decay of the average twist angle
\begin{equation}
l_\text{t} (m) \equiv -
\frac{ma}{\log \left\langle \cos \sum_{k=n}^{n+m-1} \Omega_3^{(k)} 
\right\rangle}.
\label{eq:ltm_numerics}
\end{equation}

Equations~(\ref{eq:lbm_numerics}) and (\ref{eq:ltm_numerics}) can be
compared to some analytical expressions. In the TWLC the bending
persistence length $l_\text{b}$ is the harmonic mean of the two bending
stiffnesses~\cite{lank03,esla09}:
\begin{equation}
l_\text{b} = \frac{2A_1A_2}{A_1+A_2}
\label{eq:lb_TWLC}
\end{equation}
while the twist persistence length is just twice the torsional stiffness 
(see e.g.\ Ref.~\citenum{brac14})
\begin{equation}
l_\text{t} = 2C.
\label{eq:lt_TWLC}
\end{equation}
The same quantities have been calculated for the MS model~\cite{nomi17}
\begin{equation}
l_\text{b} = 2A_1 \frac{A_2-G^2/C}{A_1+A_2-G^2/C} 
\label{eq:lb_MS}
\end{equation}
and 
\begin{equation}
l_\text{t} = 2C \left( 1 - \frac{G^2}{A_2C}\right).
\label{eq:lt_MS}
\end{equation}
From the last two expressions one recovers the TWLC limit
upon setting $G=0$.

%%%%%%%%%%%%%%%%%%%%%%%%%%%%%%%%%%%%%%%%%%%%%%%%%%%%%%%%%%%%%%%
\setlength{\tabcolsep}{7pt} % Space between columns

\begin{table*}[t!]
\caption{Elements of the stiffness matrix (expressed in nm) for
different base pairs, obtained from all-atom simulations (courtesy of
F. Lanka{\v{s}} and T.~Dr{\v{s}}ata). In order to facilitate the readout,
we have included the tilt, roll and twist nomenclature, which corresponds
to our definition of $\Omega_1$, $\Omega_2$ and $\Omega_3$, respectively.}

\centering
%  \begin{ruledtabular}
\begin{tabular}{l @{\hspace{-2pt}} l |r r r r r r r r r r |r}

&& CG &CA &TA &AG &GG &AA &GA &AT &AC &GC& average\\
	
\hline

$A_1$ & (tilt-tilt)& 47.6 & 50.6 & 44.5 & 67.3&
70.7       & 60.9 & 69.9 & 73.6 & 75.0&
70.0& 63.0\\
$A_2$ & (roll-roll)& 27.7 & 31.4 & 24.5 & 41.0&
44.4       & 42.2 & 38.7 & 45.1 & 46.1&
47.3& 38.8\\
$C$ & (twist-twist)& 32.7 & 34.0 & 57.6 & 57.9&
58.9       & 49.5 & 46.6 & 77.7 & 65.1&
51.7& 53.2\\
$G$ & (roll-twist) & 3.7  &  5.8 & 14.1 &  6.7&
7.4        &10.5  & 15.7 & 11.9 & 13.4&
13.0& 10.2 \\
$M_{12}$ & (tilt-roll)& 2.8  &  1.3 &  0.1 & -5.3&
-1.7       & 3.6  & -0.2 &  0.4 &  4.0&
-0.5& 0.4\\
$M_{13}$ & (tilt-twist) & 4.4  & -1.5 & -1.1 & -3.9&
0.9        & 6.7  &  0.0 & -0.7 & -0.6&
-0.7& 0.4
\end{tabular}

\label{table2}
%  \end{ruledtabular}
\end{table*}
%%%%%%%%%%%%%%%%%%%%%%%%%%%%%%%%%%%%%%%%%%%%%%%%%%%%%%%%%%%%%%%

Figure~\ref{fig:lb-lt} shows a comparison of the persistence lengths, as
obtained from Eq.~\eqref{eq:lbm_numerics} and \eqref{eq:ltm_numerics},
with the analytical expressions of the TWLC (Eqs.~\eqref{eq:lb_TWLC}
and \eqref{eq:lt_TWLC}) and the MS model (Eqs.~\eqref{eq:lb_MS} and
\eqref{eq:lt_MS}). There is a good overall agreement between the direct
computation of the persistence lengths and Eqs.~\eqref{eq:lb_MS} and
\eqref{eq:lt_MS} (with the plateau values of Fig.~\ref{fig:elast_param}),
for both oxDNA1 and oxDNA2.  In particular, the prediction of the
twisting persistence length is excellent in both models, whereas some
small deviations are observed for $l_\text{b}$ (smaller than 10~\%).
This suggests that there are some features of oxDNA which are not
fully captured by the ``projection'' to an inextensible elastic model,
as described by Eq.~(\ref{eq:energy_gen}). Note that $l_\text{b}$ in
oxDNA2 exhibits a damped oscillatory behaviour at short lengths $m$
with the helix periodicity, suggesting that the tangent vectors are
systematically misaligned. The value of the bending persistence length
calculated here is in agreement with previous published oxDNA1 and oxDNA2
data~\cite{ould11,mate15,snod15}.

\section{Discussion}

Owing to its chirality, DNA has been found to possess some 
remarkable mechanical properties, such as twist-bend~\cite{mark94} and 
twist-stretch coupling~\cite{mark97}. Although the latter has been 
investigated in several studies~\cite{gore06, lion07, upma08, shei09, lipf14}, 
the effect of twist-bend coupling remains to date largely unexplored.
Motivated by some recently resurgent interest~\cite{nomi17}, 
we have investigated the origin of this interaction in oxDNA,
a coarse-grained model of nucleic acids. Twist-bend coupling is a
cross-interaction between twist and bending degrees of freedom. In the
context of DNA, the existence of such an interaction was predicted
by Marko and Siggia~\cite{mark94}, who argued that twist-bend
coupling follows from the groove asymmetry, a characteristic of the
DNA molecular structure.  
% Although this prediction is more than 20
% years old, twist-bend coupling has received little attention in the
% DNA literature. The standard polymer model currently used to describe
% DNA elasticity remains the TWLC, which is characterized by a diagonal
% stiffness matrix (Eq.~(\ref{eq:energy_twlc})), with no cross-terms.

OxDNA is particularly suited to investigate the origin of twist-bend
coupling, as it comes in two different versions (oxDNA1 and oxDNA2).
The double helical grooves are symmetric in oxDNA1 and asymmetric
in oxDNA2, with widths reproducing the average B-DNA geometry. Our
simulations, sampling equilibrium conformations of both oxDNA1 and
oxDNA2, show that only the latter model has a significant twist-bend
coupling term (Fig.~\ref{fig:elast_param}). This is in agreement with
the symmetry argument by Marko and Siggia~\cite{mark94}.

The estimated twist-bend coupling coefficient from oxDNA2 is
$G=30\pm1$~nm, which agrees with the value $G=40\pm10$~nm, obtained
from fitting magnetic tweezers data~\cite{nomi17}. An earlier estimate
of $G\approx 25$~nm was obtained from the analysis of structural
correlations of DNA wrapped around histone proteins~\cite{moha05}.
It is worth noting that all-atom simulations also support the existence
of a twist-bend coupling term~\cite{lank00,lank03, drvs14}, although
those studies are restricted to short fragments ($\approx 20$~bp).
Table~\ref{table2} contains the elements of one-step stiffness matrices,
obtained by Lanka{\v{s}} et al.~\cite{lank03} from all-atom simulations.

Although the original analysis included various stretching deformations,
here we only show the rotational coordinates, while the translational
degrees of freedom are integrated out. The data in Table~\ref{table2}
refer to deformations between neighboring base pairs, hence they
are the counterparts of the $m=1$ data of Fig.~\ref{fig:elast_param}
and cannot be used as reliable estimates of asymptotic values of the
elastic parameters. Nonetheless, the averages over all possible sequence
combinations (last column of Table~\ref{table2}) show that twist-bend
coupling is much larger than the other off-diagonal terms, i.e.\ $G \gg
M_{12}, M_{13}$.

One of the most remarkable effects of twist-bend coupling in DNA
is the appearance of a novel twist length scale~\cite{nomi17}
(Eq.~(\ref{eq:lt_MS})) with an associated twist stiffness
$\kappa_\text{t} = l_\text{t}/2$, which differs from the intrinsic value
$C$. We refer to $\kappa_\text{t}$ as the renormalized
twist stiffness. In the MS and TWLC models a pure twist deformation
($\Omega_1=\Omega_2=0$, $\Omega_3 \neq 0$) has an associated intrinsic
stiffness $C$. In the presence of bending fluctuations ($\langle
\Omega_1^2 \rangle$, $\langle \Omega_2^2 \rangle > 0$), however,
the two models behave differently. While the torsional stiffness of
the TWLC remains the same, in the MS model twist deformations are
governed by a lower stiffness $\kappa_\text{t} < C$. 
% In other words, twist-bend coupling allows for the relief of twist strain by 
% means of induced bending, therefore making the DNA molecule torsionally softer.
In other words, in the presence of bending fluctuations, twist-bend
coupling makes the DNA molecule torsionally softer. From oxDNA2
simulations we estimate $\kappa_\text{t} = l_\text{t}/2\approx 83$~nm
(see Fig.~\ref{fig:lb-lt}). This is close to the value $\kappa_\text{t} =
75$~nm, recently obtained from fitting the MS model to magnetic tweezers
data~\cite{nomi17}. The above effect naturally explains~\cite{nomi17}
some reported discrepancies in the experimental determination of $C$.

Having shown that the twist-bend coupling is a relevant interaction in
DNA, one can ask in which limits and for which quantities the TWLC can
still be considered a good DNA model. Our work shows that one can map
freely fluctuating DNA onto a TWLC using $C\approx 80$~nm as twist elastic
parameter, which incorporates the effect of twist-bend coupling.  However
some care needs to be taken in the presence of a stretching force, as the
suppression of bending fluctuation will influence the twist stiffness.
At high forces DNA will then be mapped onto an effective TWLC with a
higher value of $C$.  Finally, it will be important to investigate the
effect of twist-bend coupling in cases where DNA behavior is influenced
by its mechanics as in DNA supercoiling~\cite{lepa15,fath15} or in
DNA-protein interactions~\cite{beck09,mark15}.

\section{Supplementary Material}

In the Supplementary Material the different triads are defined and
the corresponding stiffness parameters are presented. Furthermore we
elaborate on how to obtain the rotation vector $\vec{\Omega}$ from
subsequent triads.  Moreover, we explored sequence-dependent effects,
by investigating some specific sequences with oxDNA. Finally, we extended
the analysis of the main text to oxRNA.


\begin{acknowledgement}

Discussions with  F.~Kriegel, F.~Lanka{\v{s}}, J.~Lipfert, C.~Matek and
W.~Vanderlinden are gratefully acknowledged. We thank T.~Dr{\v{s}}ata
for analyzing the all-atom simulation trajectories~\cite{lank03}, from
which stiffness data in Table~\ref{table2} were obtained. We acknowledge
financial support from KU Leuven grant IDO/12/08, and from the Research
Funds Flanders (FWO Vlaanderen) grant VITO-FWO 11.59.71.7N

\end{acknowledgement}

\section*{Supporting Information for ``DNA elasticity from
coarse-grained simulations: the effect of groove asymmetry"}

%  \maketitle

This document contains additional information and results in support of
the main manuscript.

\section*{Triad Definitions}
\subsection*{Continuous Chain}

In order to describe any local deformations of an inextensible,
elastic rod, onto which DNA can be mapped, one has to introduce a local
frame of reference $\{\vec{\widehat{e}}_1(s), \vec{\widehat{e}}_2(s),
\vec{\widehat{e}}_3(s)\}$ (triad) to every point along the rod. The
deformations can thus be determined from the rotation of one triad
into the next one.  In the case of a continuous chain, the following
differential equation will hold
\begin{equation}\label{eq:DE_rotation}
\frac{\dd\vec{\widehat{e}}_\mu}{\dd s} = 
( \vec{\Omega} + \omega_0 \vec{\widehat{e}}_3 ) 
\times \vec{\widehat{e}}_\mu
\end{equation}
and this frame of reference can be unambiguously defined: $\uvec{e}_3$ may
be taken to be the tangent to the curve, $\uvec{e}_1$ pointing along the
symmetry axis of the two grooves (oriented towards the major groove) and
$\uvec{e}_2$ simply given by $\uvec{e}_2 = \uvec{e}_3 \times \uvec{e}_1$.

\subsection*{oxDNA} 

%%%%%%%%%%%%%%%%%%%%%%%%%%%%%%%%%%%%%%%%%%%%%%%%%%%%%%%%%%%%%%%%%%%%%
\begin{figure}
\centering
\includegraphics[width=0.55\linewidth]{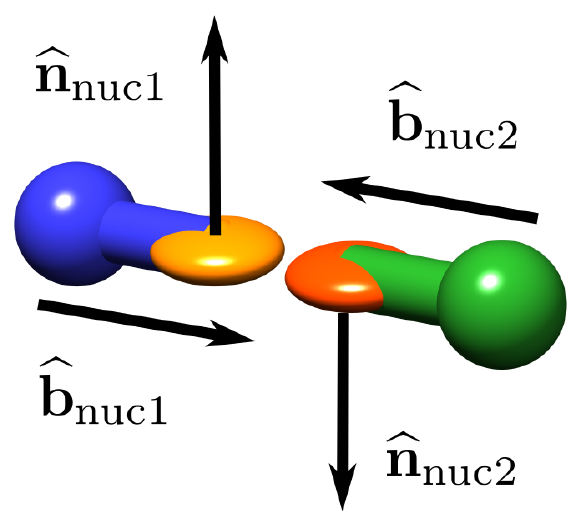}
\caption{Besides the center-of-mass position, oxDNA stores two vectors
for each nucleotide.  A unit vector~$\uvec{b}$ connects the backbone
with the base, and a normal vector~$\uvec{n}$ defines the plane of the
base. The blue and green spheres represent the backbone sites, whereas
the yellow and orange ones correspond to the base planes.}
\label{fig:supp_orientation}
\end{figure}  
%%%%%%%%%%%%%%%%%%%%%%%%%%%%%%%%%%%%%%%%%%%%%%%%%%%%%%%%%%%%%%%%%%%%%

In the discrete case of oxDNA, different triads can be defined using the
few reference points provided by the coarse-grained model.  In particular,
oxDNA consists of rigid nucleotides represented by three interactions
sites: the hydrogen-bonding, stacking and backbone sites (T. Ouldridge,
PhD Thesis, University of Oxford (2011)).  The orientation of each
nucleotide is given by a normal vector $\uvec{n}$, specifying the plane
of the base, and a vector $\uvec{b}$ pointing from the stacking site to
the hydrogen-bonding site (as in Fig.~\ref{fig:supp_orientation}). For
oxDNA1 all three sites lie on the same straight line, while in oxDNA2 the
position of the backbone site is changed, thus inducing the grooving
asymmetry (B.E. Snodin et al. J. Chem. Phys. \textbf{142}, 234901
(2015)).  Hence each base-pair comes with 2 intrinsic triads (one per
nucleotide), with the normal vectors pointing in the respective 5'-3'
direction of the strands. The interactions are designed such  that in
the minimum energy configuration the vectors $\uvec{b}_{\text{nuc1}}$
and $\uvec{b}_{\text{nuc2}}$, attached to the two nucleotides of the
same base-pair, point directly towards each other.

In what follows we present the four different choices of 
triads we have tested.

\paragraph*{Triad I.}
\label{para:triadI}

% This immidiately presents a very natural definition of the triads by the 
% exclusive use of those intrinsic triads. 
The aforementioned intrinsic nucleotide triads present a natural definition for 
the triad attached to a base pair. The base-pair normal vector can be 
constructed as the average vector of the nucleotide normal vectors
\begin{equation}
\uvec{e}_3 = \frac{\uvec{n}_{\text{nuc1}} 
- \uvec{n}_{\text{nuc2}}}{\norm{\uvec{n}_{\text{nuc1}} 
- \uvec{n}_{\text{nuc2}}}}.
\end{equation} 
The mean vector of $\uvec{b}_{\text{nuc1}}$ and $\uvec{b}_{\text{nuc2}}$ 
\begin{equation}
\uvec{y} = \frac{\uvec{b}_{\text{nuc1}}-\uvec{b}_{\text{nuc2}}}
                {\norm{\uvec{b}_{\text{nuc1}}-\uvec{b}_{\text{nuc2}}}}
\end{equation}
can be approximately identified with $\uvec{e}_2$, however in general it will 
fail to be orthogonal to $\uvec{e}_3$. This can easily be rectified by 
projecting it onto the orthogonal space of $\uvec{e}_3$ 
\begin{equation}
\vec{\widehat{e}}_2 = \frac{\uvec{y} - (\uvec{y} \cdot 
{\vec{\widehat{e}}_3) \vec{\widehat{e}}_3}}
{\norm{\uvec{y} - (\uvec{y} \cdot {\vec{\widehat{e}}_3) \vec{\widehat{e}}_3}}}.
\label{eq:parameters_e2def}
\end{equation} 
The last vector is simply given by $\uvec{e}_1 = \uvec{e}_2 \times \uvec{e}_3$.

\paragraph*{Triad II.}
\label{para:triadII}

Alternatively, $\uvec{e}_2$ can be obtained from connecting the centers
of mass $\vec{r}_{\text{nuc1}}$ and $\vec{r}_{\text{nuc2}}$ of the
two nucleotides
\begin{equation}
\uvec{y} = \frac{\vec{r}_{\text{nuc1}}-\vec{r}_{\text{nuc2}}}
                {\norm{\vec{r}_{\text{nuc1}}-\vec{r}_{\text{nuc2}}}}
\end{equation}
and the complete triad can be found in a completely analogous way as for
Triad I.  This particular choice of triad was used in the main article,
as it appeared to be the most robust (i.e.\ it yielded the smallest
correlations between consecutive $\Omega_\mu$).

\paragraph*{Triad III.}
\label{para:triadIII}
The tangent vector can also be constructed using the center of mass of the 
nucleotides. The center of mass of the i-th basepair can be defined as 
\begin{equation}
\vec{R_{\text{bp}}}(i) = \frac{\vec{r}_{\text{nuc1}}(i)+\vec{r}_{\text{nuc2}}(i)}{2}.
\end{equation}
Identifying the normalized connectors of consecutive
$\vec{R_{\text{bp}}}(i)$ with $\uvec{e}_3$ would result in a
directionally-dependent definition, therefore $\uvec{e}_3$ was chosen
as the connector between the center of masses of the previous and next
basepair
\begin{equation}
\uvec{e}_3(i) = 
\frac{\vec{R_{\text{bp}}}(i+1)-\vec{R_{\text{bp}}}(i-1)}{\norm{\vec{R_{\text{bp}
}}(i+1)-\vec{R_{\text{bp}}}(i-1)}}.
\end{equation}
The definition of the remaining triad versors is identical to the one
used for Triad II.

\paragraph*{Triad IV.}
\label{para:triadIV}
Instead of selecting one vector as the arithmetic mean and projecting
the others on its orthogonal space, one can attempt to treat them on
a more equal footing. By placing the 3 nucleotide triad vectors in the
columns of a matrix one obtains a rotation matrix
\begin{equation}
\vec{T}_\text{nuc} = [\uvec{t}_{\text{nuc}},\uvec{b}_{\text{nuc}},\uvec{n}_{\text{nuc}}] \in SO(3),
\end{equation} 
with $\uvec{t}_{\text{nuc}} = \uvec{b}_{\text{nuc}} \times
\uvec{n}_{\text{nuc}}$. The arithmetic mean $\overline{\vec{T}} =
\frac{1}{2}\left(\vec{T}_\text{nuc1} + \vec{T}_\text{nuc2}\right)$
will generally not be a rotation matrix itself, it is however possible
to orthogonally project $\overline{\vec{T}}$ onto $SO(3)$. It can be
shown that this projection is given by (M. Moakher, SIAM J. Matrix
Anal. Appl. {\bf 24}, 1 (2002))
\begin{equation}
\vec{T} = \overline{\vec{T}}\vec{U}\;\text{diag}\left(\frac{1}
{\sqrt{\Lambda_1}}\text{,}\frac{1}{\sqrt{\Lambda_2}}\text{,}\frac{s}{\sqrt{
\Lambda_3}}\right)\vec{U}^\intercal,
\end{equation}
where $\overline{\vec{T}} = \frac{1}{N}\sum_{k=1}^N T^{(k)}$,
$\Lambda_1 \geq \Lambda_2 \geq \Lambda_3 \geq 0$ are the eigenvalues
of $\vec{M} = \overline{\vec{T}}^\intercal \overline{\vec{T}}$ and the
matrix $\vec{U}$ is defined so that $\vec{U}^\intercal\vec{M}\vec{U}
= \text{diag}(\Lambda_1,\Lambda_2,\Lambda_3)$. The variable $s$
satisfies $s=1$ if $\det \overline{\vec{T}} > 0$ and $s=-1$ if $\det
\overline{\vec{T}} < 0$.

\section*{Calculation of $\Omega$}

Eq.~\eqref{eq:DE_rotation} (valid for infinitesimal rotations) can
be generalized for finite rotations.  According to Rodrigues' rotation
formula, the rotation of a vector $\vec{v}$ about an axis $\uvec{\Theta}$
by an angle $\Theta$ is given by
\begin{eqnarray}
\vec{v}_{\text{rotated}} &=& \vec{v}\cos\Theta + \left(\uvec{\Theta} \times 
\vec{v}\right) \sin\Theta + \nonumber\\
&&\uvec{\Theta}\left(\uvec{\Theta}\cdot\vec{v}\right)\left(1-\cos\Theta\right).
\end{eqnarray}
From each triad one can construct an orthogonal matrix, by placing the triad 
vectors in the columns of a $3\times3$ matrix
\begin{equation}
\vec{T}(i) = [\uvec{e}_{1}(n),\uvec{e}_{2}(n),\uvec{e}_{3}(n)] \in SO(3).
\end{equation}
This matrix is exactly the rotation matrix, transforming the canonical
frame into the frame of the respective triad. The matrix rotating
$\vec{T}(n)$ into $\vec{T}(n+1)$ with respect to the coordinate system of
the $n$-th triad is given by
\begin{equation}\label{eq:rot_fromtriads}
\vec{R} = \vec{T}^\intercal(n)\vec{T}(n+1).
\end{equation}
It is straightforward to show that in this frame the rotation matrix
$\vec{R}$ can by written in terms of the components of the rotation
vector\footnote{Note that $\Theta$ is now written in terms of the
basis of the $n$-th triad $\vec{\Theta}\DID{n} = \Theta_1\DID{n}
\uvec{e}_{1}(n) + \Theta_2\DID{n} \uvec{e}_{2}(n)+\Theta_3\DID{n}
\uvec{e}_{3}(n)$. In the remainder of this section the superscript is
omitted to enhance the readability.} $\vec{\Theta} = \left(\Theta_1\;
\Theta_2\; \Theta_3\right)^\intercal$
{\begin{onecolumn}
\begin{equation}\label{eq:rodrigues_rotationmatrix}
 \vec{R}(\vec{\Theta}) = \begin{pmatrix}
\cos\Theta + \left(\frac{\Theta_1}{\Theta}\right)^2\left(1-\cos\Theta\right) 
& \frac{\Theta_1\Theta_2}{\Theta^2}\left(1-\cos\Theta\right) - \frac{\Theta_3}{\Theta}\sin\Theta   
& \frac{\Theta_1\Theta_3}{\Theta^2}\left(1-\cos\Theta\right) + 
\frac{\Theta_2}{\Theta}\sin\Theta 
\\[4pt]
\frac{\Theta_1\Theta_2}{\Theta^2}\left(1-\cos\Theta\right) + 
\frac{\Theta_3}{\Theta}\sin\Theta   
& \cos\Theta + \left(\frac{\Theta_2}{\Theta}\right)^2\left(1-\cos\Theta\right) 
& \frac{\Theta_2\Theta_3}{\Theta^2}\left(1-\cos\Theta\right) - \frac{\Theta_1}{\Theta}\sin\Theta 

\\[4pt]
\frac{\Theta_1\Theta_3}{\Theta^2}\left(1-\cos\Theta\right) - \frac{\Theta_2}{\Theta}\sin\Theta  
& \frac{\Theta_2\Theta_3}{\Theta^2}\left(1-\cos\Theta\right) + 
\frac{\Theta_1}{\Theta}\sin\Theta   
& \cos\Theta + \left(\frac{\Theta_3}{\Theta}\right)^2\left(1-\cos\Theta\right)
               \end{pmatrix}.
\end{equation}
\end{onecolumn}}

\begin{twocolumn}
The components of $\vec{\Theta}$ can now be extracted by equating
Eqs.~\eqref{eq:rot_fromtriads} and \eqref{eq:rodrigues_rotationmatrix}
and solving for $\Theta_1$, $\Theta_2$ and $\Theta_3$. A simple way to
do this is by noticing that
\begin{equation}\label{eq:trace_of_rotmat}
\text{tr} (\vec{R}) = 1 + 2\cos\Theta.
\end{equation}
Moreover, one can also verify that the following relation holds
\begin{equation}\label{eq:Omega_from_rodrigues}
\vec{\Theta}  = \frac{\Theta}{2\sin\Theta} 
\begin{pmatrix}
R_{32} - R_{23} \\[4pt]
R_{13} - R_{31} \\[4pt]
R_{21} - R_{12}
\end{pmatrix}.
\end{equation}
Note that the sign ambiguity presented in Eq.~\eqref{eq:trace_of_rotmat} is 
completely inconsequential for Eq.~\eqref{eq:Omega_from_rodrigues}. 

We define the deformation parameters $\Omega_\mu$ as the deviations
of the components of $\vec{\Theta}/a$ from their respective mean value
\begin{equation}
a\Omega_\mu \equiv \Theta_\mu - \la \Theta_\mu \ra.
\end{equation}
For an ideal triad definition, the mean values of $\Theta_1$ and
$\Theta_2$ are expected to be zero, while $\la\Theta_3\ra /a$ should be
equal to the intrinsic twist $\omega_0$. In the case of oxDNA2, the mean
value of $\Theta_2$ is in fact distinctly non-zero (about $2.6^\circ$
for Triad II and very similar for the other triads), resulting in the
oscillatory behavior of the persistence length shown in Fig.~5 of the
main text.  All triad definitions consistently yield $\la \Theta_3
\ra \approx 34.8 ^\circ$ and $\la \Theta_3 \ra \approx 34.1 ^\circ$
for oxDNA1 and oxDNA2 respectively.

%%%%%%%%%%%%%%%%%%%%%%%%%%%%%%%%%%%%%%%%%%%%%%%%%%%%%%%%%%%%%%%%%%%%%
\begin{figure}[H]
\hspace*{-0.55cm}\includegraphics[width=1.00\linewidth]{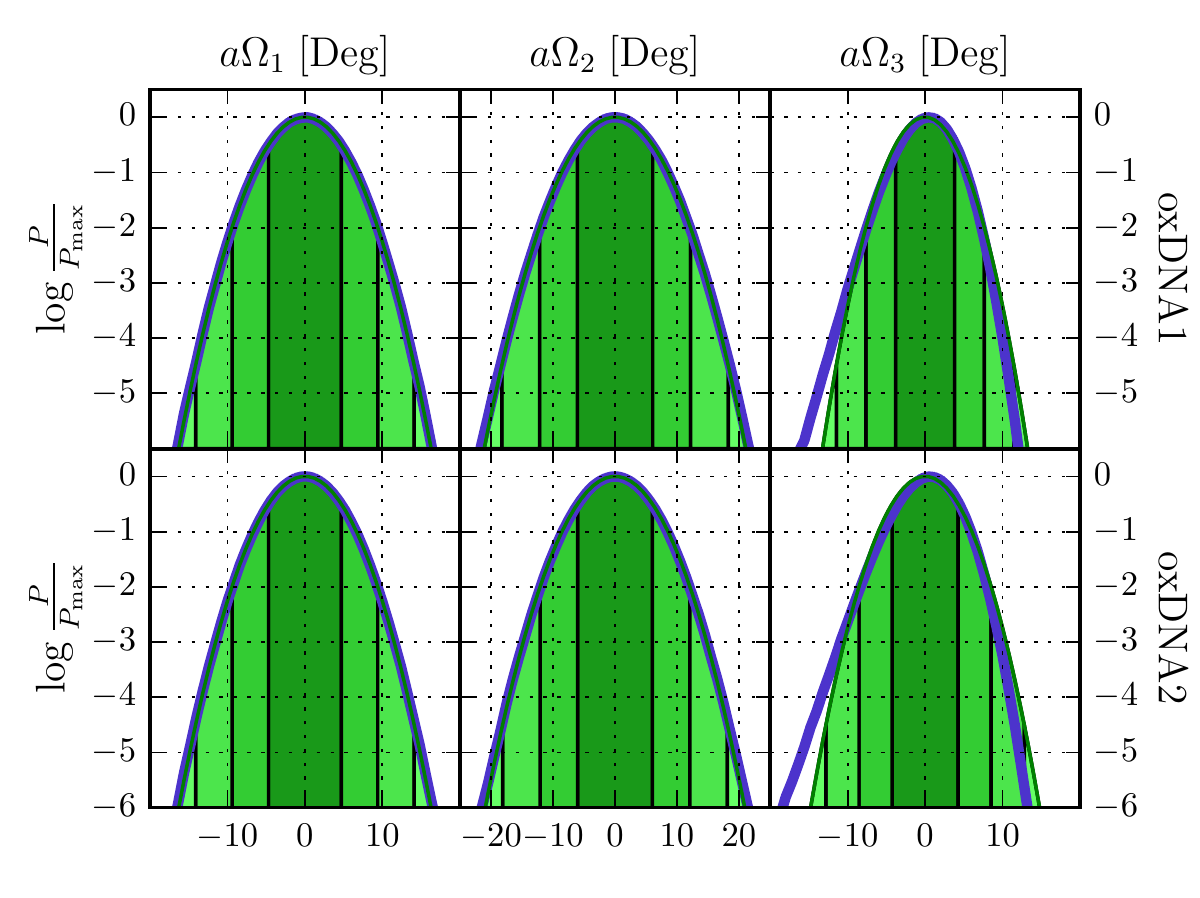}
\caption{Probability distribution of $\Omega_\mu$ for oxDNA1 and oxDNA2 in 
logscale, using Triad II. The distributions coincide very well with Gaussian 
distributions, validating the use of a quadratic form for the free energy. 
Analogous distributions are found for the other choices of triads.}
\label{fig:ProbDist}
\end{figure}
%%%%%%%%%%%%%%%%%%%%%%%%%%%%%%%%%%%%%%%%%%%%%%%%%%%%%%%%%%%%%%%%%%%%%

\section*{Distributions of $\Omega$'s}

The approximation of the free energy by a quadratic form
\begin{eqnarray}
\beta E &=& \frac{a}{2} \sum_{n=1}^N 
\left(\sum_{\mu,\nu=1}^3 
\Omega_\mu\DID{n}
M_{\mu\nu}
\Omega_\nu\DID{n}\right)
\label{eq:SUPP:elastic_energy_discrete}
\end{eqnarray}
implicitly assumes that the deformation parameters follow a Gaussian
distribution. Figure~\ref{fig:ProbDist} shows the distributions of
$\Omega_1$, $\Omega_2$ and $\Omega_3$ (blue lines) as obtained from
equilibrium simulations. For a clear comparison, the distributions are
shown in logarithmic scale.  The fitted Gaussian curves (green lines)
indicate that the quadratic approximation is excellent for $\Omega_1$
and $\Omega_2$, while some small deviations are observed in the
distributions of $\Omega_3$ (noticeable for angles larger than $15$
degrees). The distributions of $\Omega_3$ are slightly asymmetric, which
is a consequence of the intrinsic twist $\omega_0$ (different response
of DNA to under- and over-twisting).

\section*{Stiffness parameters for alternative triads definitions}

The extracted stiffness parameters for the 4 different
triads are summarized in Fig.~\ref{fig:mCoupling4Triads} and
Table~\ref{table:alltriads}. The plateau values (large $m$) are quite
consistent among the different triad definitions, with the exception of
$A_1$ and $A_2$ obtained from Triad III.  On the other hand, the values
obtained for $m=1$ are significantly more diverse.

%%%%%%%%%%%%%%%%%%%%%%%%%%%%%%%%%%%%%%%%%%%%%%%%%%%%%%%%%%%%%%%%%%%%
\begin{figure*}[t]
\centering\includegraphics[width=\linewidth]{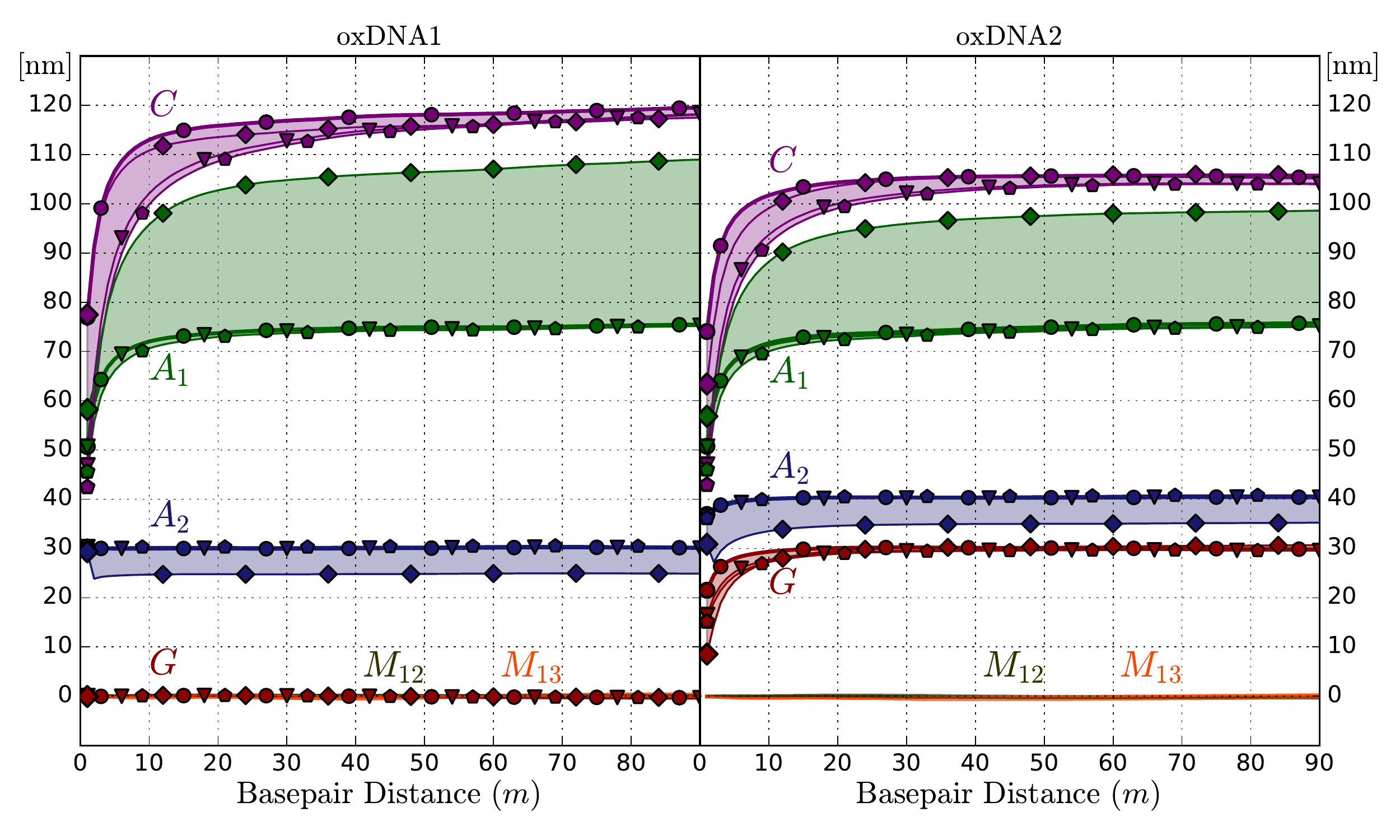}
\caption{Elements of the $m$-step stiffness matrix as a function of the
base-pair distance $m$, extracted from 4 different triad definitions.
Data based on Triad I, II, III and IV are shown with triangles, circles,
diamonds and pentagons, respectively. Note that the spread in the plateau
values of $G$ and $C$ is remarkably small, despite the large differences
at $m=1$. For the bending stiffness parameters $A_1$ and $A_2$, Triads
I, II and IV practically yield the same plateau values, while Triad
III tends to give quite different values. This is probably due to the
fundamentally different definition of the tangent vector in Triad III.}
\label{fig:mCoupling4Triads}
\end{figure*}
%%%%%%%%%%%%%%%%%%%%%%%%%%%%%%%%%%%%%%%%%%%%%%%%%%%%%%%%%%%%%%%%%%

%%%%%%%%%%%%%%%%%%%%%%%%%%%%%%%%%%%%%%%%%%%%%%%%%%%%%%%%%%%%%%%
\setlength{\tabcolsep}{7pt} % Space between columns

\begin{table*}[h!]
\caption{Values of the stiffness coefficients (expressed in nm) for oxDNA1 and 
oxDNA2 for different Triad definitions. The values given here correspond to the 
plateau values of Fig.~\ref{fig:mCoupling4Triads}.}
\centering
%  \begin{ruledtabular}
\begin{tabular}{ l | c c c r | c c c r}

%& \multicolumn{4}{c}{Team sheet} & \multicolumn{4}{c}{Team sheet}\\

&\multicolumn{4}{c|}{oxDNA1} & 
\multicolumn{4}{c}{oxDNA2} \\

%& & oxDNA1 & & & & oxDNA2 & & \\

& $A_1$ & $A_2$ & C & \multicolumn{1}{c|}{G} & $A_1$ & $A_2$ & C & \multicolumn{1}{c}{G}\\
	
\hline

Triad I   & 76 & 30 & 120 & 0.1 & 76 & 40 & 105 & 29.8 \\
Triad II  & 75 & 30 & 118 & 0.2 & 75 & 40 & 104 & 29.6 \\ 
Triad III & 109& 25 & 118 & -0.3& 99 & 35 & 106 & 30.7 \\ 
Triad IV  & 75 & 30 & 118 & 0.1 & 75 & 41 & 104 & 29.6

\end{tabular}
\label{table:alltriads}
%  \end{ruledtabular}
\end{table*}
%%%%%%%%%%%%%%%%%%%%%%%%%%%%%%%%%%%%%%%%%%%%%%%%%%%%%%%%%%%%%%%

\section*{Sequence Dependence}

So far we have ignored any sequence-dependent effects in oxDNA
by considering average base-pair interaction coefficients. This is
expected to be a valid approximation for typical and sufficiently long
DNA sequences (i.e.\ consisting of hundreds of base pairs), for which
such effects are averaged out.

In order to explore the impact of sequence-dependent interactions, we
have repeated the analysis of the main text for some special choices
of sequences. The results, for both oxDNA1 and oxDNA2, are summarized
in Fig.~\ref{fig:SeqDep} and Table~\ref{table:SeqDep}. All parameters
exhibit relatively small variations (within 15~\%), and $G$ remains
significantly non-zero in all cases for oxDNA2.

%%%%%%%%%%%%%%%%%%%%%%%%%%%%%%%%%%%%%%%%%%%%%%%%%%%%%%%%%%%%%%%%%%%%
\begin{figure*}[t]
\centering\includegraphics[width=\linewidth]{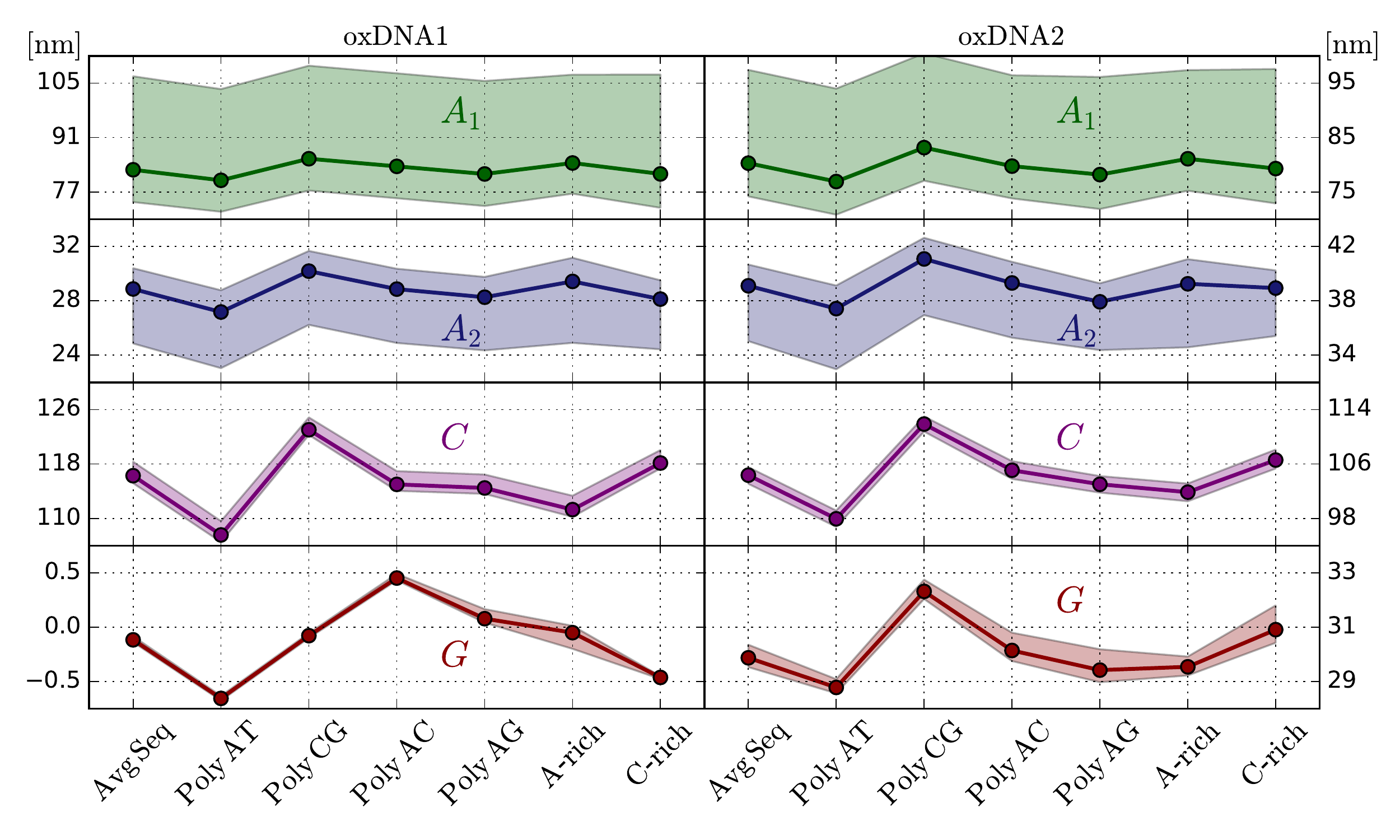}
\caption{Plateau values of the elastic constants, calculated in the same
way as described in the main text, for some specific sequences of dsDNA.
The indicated values are the mean values over the 4 different triad
definitions, while the shaded region indicates the range of observed
values.  From these plots it is clear that the interaction parameters
are only weakly sequence-dependent.  Furthermore, we see again that only
for oxDNA2 the coupling term $G$ is significantly non-zero.  A-rich and
C-rich indicate sequences which contained approximately 83\% of A and
C respectively. The `Poly'-sequences consisted out of repetitions of
two bases.  Numerical values can be found in Table~\ref{table:SeqDep}.
For comparison we included the values found with the averaged sequence
parameters.}
\label{fig:SeqDep}
\end{figure*}
%%%%%%%%%%%%%%%%%%%%%%%%%%%%%%%%%%%%%%%%%%%%%%%%%%%%%%%%%%%%%%%%%%

\begin{table*}[h!]
\caption{The elastic constants for some specific sequences of
dsDNA.  These are the numerical values of the quantities plotted in
Fig.~\ref{fig:SeqDep}, averaged over all four triad definitions. More
information can be found in the caption of Fig.~\ref{fig:SeqDep}}
\centering
%  \begin{ruledtabular}
\begin{tabular}{ l | c c c c | c c c c}

% \multicolumn{2}{c}{} & \multicolumn{2}{c}{oxDNA1} & & & 
% \multicolumn{2}{c}{oxDNA2} & \multicolumn{1}{c}{}\\
& \multicolumn{4}{c|}{oxDNA1} & 
\multicolumn{4}{c}{oxDNA2} \\

& $A_1$ & $A_2$ & C & \multicolumn{1}{c|}{G} & $A_1$ & $A_2$ & C & \multicolumn{1}{c}{G}\\	
\hline
AvgSeq   & 82.7 & 28.9 & 116.3 & -0.12 & 80.3 & 39.1 & 104.4 & 29.9 \\
Poly AT  & 80.0 & 27.2 & 107.6 & -0.65 & 76.9 & 37.4 & \textcolor{white}{0}97.9 
& 28.8 \\ 
Poly CG  & 85.6 & 30.2 & 123.0 & -0.08 & 83.1 & 41.1 & 111.9 & 32.3 \\
Poly AC  & 83.6 & 28.9 & 115.0 & \hphantom{-}0.45 & 79.8 & 39.3 & 
105.1 & 30.2 \\ 
Poly AG  & 81.6 & 28.3 & 114.5 & \hphantom{-}0.08 & 78.2 & 37.9 & 103.0 
& 29.4 \\
A-rich   & 84.5 & 29.4 & 111.3 & -0.05 & 81.1 & 39.3 & 101.9 & 29.5 \\
C-rich   & 81.6 & 28.1 & 118.1 & -0.46 & 79.3 & 38.9 & 106.6 & 30.9

\end{tabular}
\label{table:SeqDep}
%  \end{ruledtabular}
\end{table*}
%%%%%%%%%%%%%%%%%%%%%%%%%%%%%%%%%%%%%%%%%%%%%%%%%%%%%%%%%%%%%%%

\section*{oxRNA}

Finally one could wonder about the magnitude of twist-bend coupling for dsRNA, 
a double-helical molecule with major and minor grooves. However, this 
double helix is in A-form, which differs from the B-form dsDNA studied in the 
main text. One of the differences is that the A-form has larger grooves, but 
there are more structural differences between the two. This makes the effect of 
the larger grooves on the magnitude of the twist-bend coupling hard to predict. 
Here we only confirmed that symmetry breaking results in a non-zero coupling, 
but it is not a priori clear which factors or structural parameters influence 
its magnitude.

To address this question more carefully, we again resort to computer
simulations. This could be done quite easily, since the authors
of oxDNA also provide a simulation code for RNA, called oxRNA. In
Fig.~\ref{fig:oxRNA} the interaction parameters from these simulations
are presented. It is important to note here that both oxRNA1 and oxRNA2
have major and minor grooves, and the difference between the two is in
modelling of electrostatic effects. For both models it is clear that $G$
is manifestly non-zero, while the other two off-diagonal terms, $M_{12}$
and $M_{13}$, lie very close to zero. This is a signature of the existence
of major and minor grooves. The value of $G$ lies around $10$~nm, which
is smaller than the one found for oxDNA. In Fig.~\ref{fig:oxRNAlblt}
the bending and twisting persistence length of RNA ($l_\text{b}$
and $l_\text{t}$, respectively) are shown. Although the magnitude of
$l_\text{b}$ and $l_\text{t}/2$ (34~nm and 73~nm for oxRNA2, respectively)
are lower than the experimentally determined ones (J.\ Lipfert et al. PNAS
\textbf{111}, 15408 (2014)), they are in line with previous estimates
in oxRNA (C.\ Matek et al. J. Chem. Phys. \textbf{143}, 243122 (2015)).
\end{twocolumn}

%%%%%%%%%%%%%%%%%%%%%%%%%%%%%%%%%%%%%%%%%%%%%%%%%%%%%%%%%%%%%%%%%%%%
\begin{figure*}[t]
\centering\includegraphics[width=\linewidth]{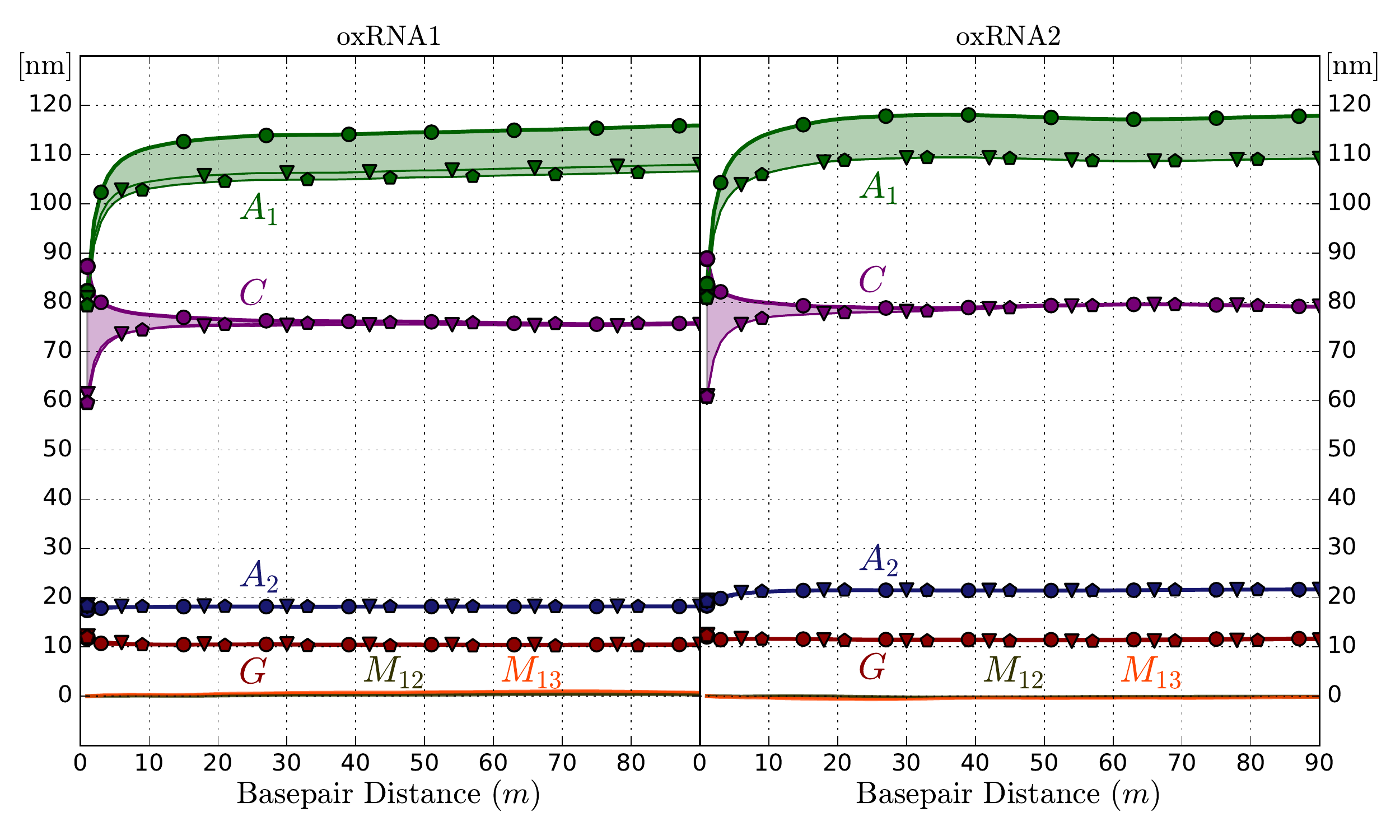}
\caption{The elastic constants obtained from oxRNA, similar to
Fig.~\ref{fig:mCoupling4Triads}. The simulation and extraction scheme was
the same as in the main text. dsRNA has major and minor grooves, and again
we observe that $G$ is non-zero, while the other two mixed interaction
parameters ($M_{12}$ and $M_{23}$) are very close to zero. It is important
to note that both oxRNA1 and oxRNA2 have major and minor grooves, but
have different implementations of the electrostatic interactions. Note
that, to enhance the readability of the plot, we have omitted the data
from Triad III, as they yielded significantly different plateau values
($A_1 = 103$~nm, $A_2 = 45$~nm, $C = 50$~nm and $G = 31$~nm, but still
$M_{12},M_{13} \approx 0$, for oxRNA2).  This is likely due to the tilted
base-pair planes with respect to the helical axis of the A-form helix,
which affects this particular choice of the tangents.}
\label{fig:oxRNA}
\end{figure*}
%%%%%%%%%%%%%%%%%%%%%%%%%%%%%%%%%%%%%%%%%%%%%%%%%%%%%%%%%%%%%%%%%%

%%%%%%%%%%%%%%%%%%%%%%%%%%%%%%%%%%%%%%%%%%%%%%%%%%%%%%%%%%%%%%%%%%%%
\begin{figure*}[t]
\centering\includegraphics[width=\linewidth]{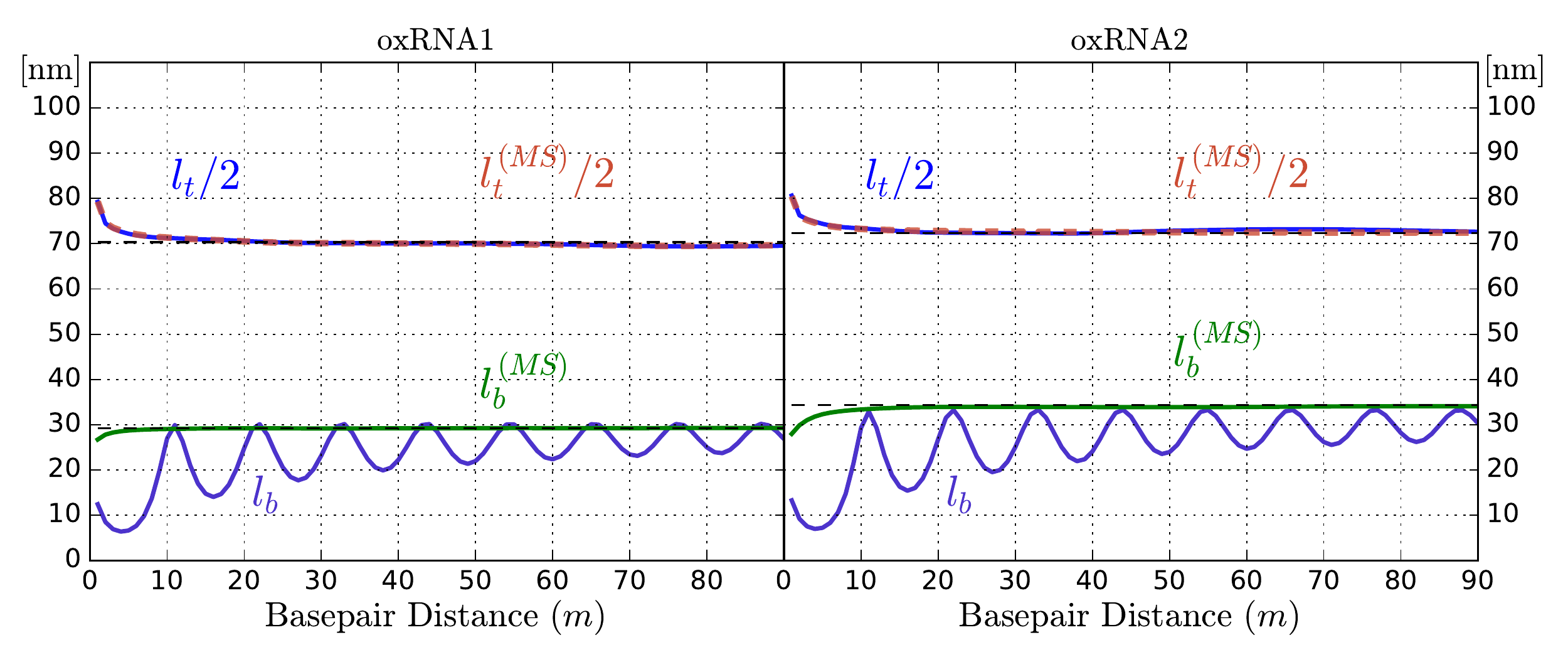}
\caption{The bending and twisting persistence length of dsRNA, similar
to Fig.~5 of the main text, as obtained from oxRNA. The values agree
well with the ones reported for oxRNA.}
\label{fig:oxRNAlblt}
\end{figure*}
%%%%%%%%%%%%%%%%%%%%%%%%%%%%%%%%%%%%%%%%%%%%%%%%%%%%%%%%%%%%%%%%%%

\end{document}